\documentstyle{article}
\author{I.G. Korepanov}
\title{Hidden Symmetries in the 6-Vertex Model of Statistical Physics}
\date{\hfill\normalsize\it Dedicated to L.D.~Faddeev on his
      60$^{\,\hbox{\scriptsize\it th}}$ birthday}
\newcommand{\be}{\begin{equation}}
\newcommand{\ee}{\end{equation}}
\newcommand{\ba}{\begin{array}}
\newcommand{\ea}{\end{array}}
\newtheorem{theorem}{Theorem}[section]
\newtheorem{lemma}{Lemma}[section]
\begin{document}
\maketitle
\smallskip
\begin{abstract}
The transfer matrix of the 6-vertex model of two-dimensional statistical
physics commutes with many (more complicated) transfer matrices, but these
latter, generally, do not commute between each other. The studying of their
action in the eigenspaces of the 6-vertex model transfer matrix
becomes possible due to a ``multiplicative property'' of the {\em vacuum
curves\/} of $\cal L$-operators from which transfer matrices are built.
This approach allowed, in particular, to discover for the first time the fact
that the dimensions of abovementioned eigenspaces must be multiples of
(big enough) degrees of the number 2.
\end{abstract}

Since the discovery in 1931 of the famous Bethe anzatz~\cite{Bethe}
for the eigenvectors of one-dimensional quantum Heisenberg magnetic model
Hamiltonian, a lot of papers were devoted to studying the properties and
generalizations of that ansatz. Now Bethe ansatz is usually considered in the
framework of quantum inverse problem method~\cite{FT} which has united in a
natural way the main achievements of one-dimensional quantum field theory and
two-dimensional statistical physics, and linked the ideas in these fields to
the exactly solvable nonlinear equations of classical mathematical physics
(``soliton equations''). Progress was achieved not only in the classification
of eigenvectors in ``thermodynamic limit'' (infinite length of the chain),
but also for the chain of finite length (see e.g.~\cite{FT-isotrop}, where
classification of eigenvectors is presented for an isotropic
magnetic---``XXX model''. Nevertheless, now still, in this paper author's
opinion, there are many undiscovered mysteries in the Bethe ansatz.

There is, however, a case in which full solution of the eigenvector problem
is not difficult---the case when the model can be reduced to ``free
fermions''. For 6-vertex model studied in this paper (and for the ``XXZ
magnetic model'' connected with it) this means that the ``coupling constant''
$\eta$ equals $\pi/4$. Baxter~\cite{Bax-book} surmised that the next
simplest cases will appear when $\eta$ equals other rational multiples of
$\pi$. Peculiar properties of such $\eta$ values became clearly seen when
exactly for those values new solutions of the Yang--Baxter equation were
constructed---the $\cal L$-operators associated with 6-vertex model
$R$-operators~\cite{IK} (see also formulae~(\ref{6}--\ref{11}) below). Next,
the author of this paper applied to those $\cal L$-operators the idea of
{\em vacuum curve}. Vacuum curves (see~\cite{Krich} about them) of such
$\cal L$-operators turned out to have a very simple
form~\cite{Kor-vac,Kor-vac-LOMI}. This together with their
multiplication properties (see Subsection~\ref{sec1} below) allowed the
author to discover for the first time that the spectrum of 6-vertex model
transfer matrix (and XXZ model Hamiltonian) is highly degenerate---the
multiplicities of degeneracy grow, roughly speaking, as $2^{{\rm const}
\cdot N}$ with $N\rightarrow\infty$, where $N$ is the chain length. The
present paper concludes the series of two papers (the first one
was~\cite{Kor-vac-LOMI}) where the author's approach to the solutions
of Yang--Baxter equation associated with algebraic curves of genus $g>1$, and
to the problem of hidden symmetries of 6-vertex model transfer matrix, is
described.

\section{The group of matrices of $\cal L$-operator vacuum curve
coefficients}\label{realsec1}
\subsection{}\label{sec1}
Consider the Yang--Baxter equation
\be R(\lambda-\mu){\cal L}(\lambda){\cal L}(\mu)={\cal L}(\mu){\cal L}
(\lambda)R(\lambda-\mu),\label{1}\ee
where $R(\lambda-\mu)$ is the $R$-matrix of 6-vertex model of two-dimensional
statistical physics~\cite{FT}. There exists, firstly, the following solution
of~(\ref{1}):
\be{\cal L}(\lambda)=\left(\ba{cccc}\sin(\lambda+\eta)&&&\\
&\sin(\lambda-\eta)&\sin2\eta&\\&\sin2\eta&\sin(\lambda-\eta)&\\
&&&\sin(\lambda+\eta)\ea\right).\label{2}\ee
A number of other solutions of~(\ref{1}) can be constructed through the
multiplication procedure~\cite{KS,KRS}. There are also ``trivial'' solutions
of~(\ref{1})---the constant $\cal L$-operators with one-dimensional quantum
space
\be{\cal L}(\lambda)=\left(\ba{cc}a_0&0\\0&d_0\ea\right)\quad\mbox{and}\quad
{\cal L}(\lambda)=\left(\ba{cc}0&b_0\\c_0&0\ea\right).\label{3}\ee

In case of a ``generic'' parameter $\eta$, all the known solutions
of~(\ref{1}) are obtained from the abovementioned ones by the following
operations: a)~construction of (inhomogeneous) monodromy matrices
\be{\cal L}(\lambda)=\stackrel{
\makebox(0,0)[lb]{\put(0,0){\vector(-3,-2){0}}}%
\mbox{\normalsize\hspace{0.03em}\raisebox{-0.3ex}{$\frown$}}
}{
\prod_{i=1}^M
}
{\cal L}^{(i)}(\lambda_i+\lambda),\label{4}
\ee
where ${\cal L}^{(i)}(\lambda)$ are arbitrary solutions of~(\ref{1}),
$\lambda_i$ are constants; b)~taking a direct sum in quantum spaces (with the
same auxiliary space)
\be{\cal L}(\lambda)=\bigoplus_{i=1}^K{\cal L}^{(i)}(\lambda_i+\lambda),
\label{5}\ee
and c)~restriction to an invariant subspace in the quantum space (if such
subspace exists; to be exact, the operator $\cal L$ is restricted to the
tensor product of auxiliary space by the invariant subspace of quantum space)
or taking the corresponding factor operator.

The situation is much more interesting if the parameter $\eta$ in~(\ref{2})
is commensurable with $\pi$. Let
$$\frac{\eta}{\pi}=\frac{m}{n}, $$
with relatively prime $m$ and $n$. Then to the solutions of~(\ref{1})
one must add the $\cal L$-matrices
\be{\cal L}(\lambda)=\left(\ba{cc}{\cal A}(\lambda)&{\cal B}(\lambda)\\
{\cal C}(\lambda)&{\cal D}(\lambda)\ea\right),\label{6}\ee
where ${\cal A}(\lambda),\ldots{\cal D}(\lambda)$ of size $n\times n$ are of
the form (with zeros at the blank spaces):
\begin{flushleft} ${\cal A}(\lambda)=$ \nopagebreak \end{flushleft}
\be= a \left( \! \ba{cccc}\! \sin(\lambda\!+\!\rho\!+\!(n\!-\!1)\eta)
\!\!\!\! &
&& \\ &\!\!\!\! \sin(\lambda\!+\!\rho\!+\!(n\!-\!3)\eta)\!\!\!\!
 && \\ &&
\ddots& \\ &&& \!\!\!\! \sin(\lambda\!+\!\rho\!+\!(1\!-\!n)
\eta) \! \ea \! \right)\!, \label{7}\ee
\medskip
\be {\cal B}(\lambda)=\left(\ba{cccc}0&&&b_{1n}\\
b_{21}&0&&\\ &\ddots&\ddots& \\ &&b_{n,n-1}&0
\ea\right),\label{8}\ee
\be {\cal C}(\lambda)=\left(\ba{cccc}0&c_{12}&&\\
&0&\ddots&\\ &&\ddots&c_{n-1,n}\\ c_{n1}&&&0
\ea\right),\label{9}\ee
\begin{flushleft} ${\cal D}(\lambda)=$ \nopagebreak \end{flushleft}
\be= d \left( \! \ba{cccc}\! \sin(\lambda\!+\!\sigma\!+\!(1\!-\!n)
\eta)\!\!\!\! &
&& \\ &\!\!\!\! \sin(\lambda\!+\!\sigma\!+\!(3\!-\!n)\eta)\!\!\!\!
& & \\ &&
\ddots& \\ &&&\!\!\!\! \sin(\lambda\!+\!\sigma\!+\!(n\!-\!1)
\eta) \! \ea \! \right)\!. \label{10}\ee
\medskip
Here $a, d, \rho, \sigma$ and all entries in matrices ${\cal B}(\lambda)$ and
${\cal C}(\lambda)$ are constants, with the following relation satisfied
(subtraction in indices is understood ${\rm mod}\;n$):
\be b_{k,k-1}c_{k-1,k}=\Delta+\frac{ad}{2}\cos\bigl(\rho-\sigma+2\eta(n-2k)
\bigr),
\label{11}\ee
$k = 1, \ldots,n;\quad\Delta$ is a constant, too.

$\cal L$-matrices of the form~(\ref{6}--\ref{11}) are interesting because
they have no {\em generating vector\/} in their quantum space
(see~\cite{IK}), i.e.\ no vector annulated by $\cal C(\lambda)$ for all
$\lambda$. Instead, they have {\em vacuum vectors\/} in the sense
of~\cite{Krich}. This leads to the important role of the vacuum curve
$\Gamma_{\cal L}(\lambda)$~\cite{Krich} of operator
${\cal L}(\lambda)$---an algebraic curve in ${\rm C}^2$ given by equation
$$\det\bigl(u{\cal A}(\lambda)+{\cal B}(\lambda)-uv{\cal C}(\lambda)-
v{\cal D}(\lambda)\bigr)=0.$$

The explicit form of $\Gamma_{\cal L}(\lambda)$ for all cases we are
interested in has been calculated in~\cite{Kor-vac}, see
also~\cite{Kor-vac-LOMI} (all the results of~\cite{Kor-vac,Kor-vac-LOMI}
that are of interest to us here are easily carried over to the general case,
with no restriction ${\cal C}(\lambda)={\cal B}(\lambda)^T$
of~\cite{Kor-vac,Kor-vac-LOMI} on the operators~(\ref{6}--\ref{11})).
The easiest case is that of the odd $n$, so we will assume this oddness up to
Subsection~\ref{sec5}.

\begin{theorem}\label{t1}{\rm\cite{Kor-vac,Kor-vac-LOMI}}.
The vacuum curve
$\Gamma_{\cal L}(\lambda)$ of an $\cal L$-operator of the
form (\ref{6}--\ref{11}) is given by equation
\be v^n=\frac{\alpha(\lambda)u^n+\beta(\lambda)}{\gamma(\lambda)u^n+
\delta(\lambda)},\label{12}\ee
where $\alpha(\lambda)=\det{\cal A}(\lambda),\ldots,\delta(\lambda)=
\det{\cal D}(\lambda).$
\end{theorem}

Let us associate with an $\cal L$-matrix of the form~(\ref{6}--\ref{11})
a matrix
\be M_{\cal L}(\lambda)=\left(\ba{cc}\alpha(\lambda)&\beta(\lambda)\\
\gamma(\lambda)&\delta(\lambda)\ea\right).\label{13}\ee
It is natural to regard $M_{\cal L}(\lambda)$ as determined up to a
meromorphic scalar factor $g(\lambda)$. Below in this paper we always assume
that $\det M_{\cal L}(\lambda)\not\equiv0$.

\begin{theorem}\label{t2} The vacuum curve $\Gamma_{\cal L}(\lambda)$ of the
monodromy matrix~(\ref{4}) composed of $\cal L$-matrices of the
form~(\ref{6}--\ref{11}) has the form
$$\left(v^n-\frac{\alpha(\lambda)u^n+\beta(\lambda)}%
{\gamma(\lambda)u^n+\delta(\lambda)}\right)^K=0,$$
where $K$ is a positive integer, and
\be\left(\ba{cc}\alpha(\lambda)&\beta(\lambda)\\
\gamma(\lambda)&\delta(\lambda)\ea\right)=
\stackrel{
\makebox(0,0)[lb]{\put(0,0){\vector(-3,-2){0}}}%
\mbox{\normalsize\hspace{0.03em}\raisebox{-0.3ex}{$\frown$}}
}{
\prod_{i=1}^M
}
M_{{\cal L}^{(i)}}
(\lambda_i+\lambda).\label{14}\ee
\end{theorem}

{\it Proof\/} follows easily from Theorem~\ref{t1} and the results of
papers~\cite{Krich,Kor-vac,Kor-vac-LOMI}.

Theorem~\ref{t2} prompts one to associate to a monodromy matrix
$\cal L(\lambda)$ as well the matrix~(\ref{14}) of its coefficients.
In this case, the condition
\be(u,v)\in\Gamma_{\cal L}(\lambda)\Rightarrow v^n=
\frac{\alpha(\lambda)u^n+\beta(\lambda)}%
{\gamma(\lambda)u^n+\delta(\lambda)}\label{15}\ee
holds, while we are paying no attention to the fact that $\Gamma_{\cal L}$
may consist of several identical components. Let us, then, associate to the
6-vertex model $\cal L$-matrix~(\ref{2}) and its multiplied versions the
identity matrix $M_{\cal L}(\lambda)$, and to the matrices~(\ref{3})
$$M_{\cal L}(\lambda)=\left(\ba{cc}a_0^n&0\\0&d_0^n\ea\right)\quad
\mbox{and}\quad M_{\cal L}(\lambda)=\left(\ba{cc}0&b_0^n\\c_0^n&0\ea\right)$$
respectively. Now allow ourselves to include in a monodromy matrix~(\ref{4})
the $\cal L$-matrices mentioned in this paragraph as well as
$\cal L$-matrices~(\ref{6}--\ref{11}). Using the results
 from~\cite{Kor-vac,Kor-vac-LOMI} we find that to such a monodromy matrix the
matrix
$$\left(\ba{cc}\alpha(\lambda)&\beta(\lambda)\\
\gamma(\lambda)&\delta(\lambda)\ea\right)$$
obtained from relation~(\ref{14}) is associated in the sense of~(\ref{15}),
as before.

Note that matrices $M_{\cal L}(\lambda)$ are {\em periodic with period\/}
$\pi/n$.

\subsection{}\label{sec2}

In papers~\cite{Kor-vac,Kor-vac-LOMI} an involution ${\cal L}
(\lambda)\mapsto\hat{\cal L}(\lambda)$ has been introduced that maps an
$\cal L$-matrix of the form~(\ref{6}--\ref{11}) into such a matrix
$\hat{\cal L}(\lambda)$ that the vacuum curve of the monodromy matrix
${\cal L}(\lambda)\hat{\cal L}(\lambda)$ has an identity matrix of
coefficients $M_{{\cal L}\hat{\cal L}}(\lambda)$. Here we will slightly
change the definition of this involution (without changing the vacuum curve
of $\hat{\cal L}(\lambda)$) and extend it to other ${\cal L}(\lambda)$ as
follows: for any $\cal L$-operator~(\ref{s2}), with the only condition on
${\cal A}(\lambda),\ldots,{\cal D}(\lambda)$ that they satisfy the 6-vertex
model commutation relations, introduce $\hat{\cal L}(\lambda)$ by the formula
\be\hat{\cal L}(\lambda)=\left(\ba{cc}{\cal D}(\lambda)^T&-{\cal B}
(\lambda)^T\\
-{\cal C}(\lambda)^T&{\cal A}(\lambda)^T\ea\right).\label{s3a}\ee

\subsection{}\label{sec3}

Now let us change the roles of quantum and auxiliary spaces of monodromy
matrices ${\cal L}(\lambda)$ introduced in the end of
Subsection~\ref{sec1} and consider for a given ${\cal L}(\lambda)$ an
inhomogeneous transfer matrix
\be T(\lambda)={\rm Tr}\stackrel{
\makebox(0,0)[lb]{\put(0,0){\vector(-3,-2){0}}}%
\mbox{\normalsize\hspace{0.03em}\raisebox{-0.3ex}{$\frown$}}
}{
\prod_{i=1}^N
}
{\cal L}(\mu_i+\lambda),\label{16}\ee
$\mu_i$ being fixed numbers, which acts in a $2^N$-dimensional linear space
$H$, the tensor product of {\em auxiliary}, from the viewpoint of
equation~(\ref{1}), spaces. It is shown in the
papers~\cite{Kor-vac,Kor-vac-LOMI} that if an identity matrix
$M_{\cal L}(\lambda)$ corresponds in the sense of~(\ref{15}) to a matrix
${\cal L}(\lambda)$ then $T(\lambda)$ commutes with the analogous transfer
matrix built up of any other matrix ${\cal L}(\lambda)$. Guided by this fact,
let us study the action of transfer matrices of the form~(\ref{16}) in a
space $H_w\subset H$, an eigenspace for all the transfer matrices
corresponding to ${\cal L}(\lambda)$'s with the identity matrix
$M_{\cal L}(\lambda)\equiv\left(\ba{cc}1&0\\0&1\ea\right)$.

\begin{theorem}\label{t3} Let monodromy matrices ${\cal L}_1(\lambda)$ and
${\cal L}_2(\lambda)$ have the same vacuum curve $\Gamma(\lambda)$ (for all
$\lambda$). Let the restrictions to $H_w$ of transfer matrices corresponding
to them according to~(\ref{16}) be non-degenerate in $\lambda=0$:
$$\det\left.\vphantom{\hat T}T_1(0)\right|_{H_w}\not=0,\quad \det\left.
\vphantom{\hat T}T_2(0)\right|_{H_w}\not=0.$$
Let, finally, exist a monodromy matrix ${\cal L}_3(\lambda)$ such that
${\cal L}_1(\lambda){\cal L}_3(\lambda)$ has an identity matrix of vacuum
curve coefficients and the transfer matrix built up of ${\cal L}_3(\lambda)$
also satisfies
$$\det\left.\vphantom{\hat T}T_3(0)\right|_{H_w}\not=0. $$

Then the equality
\be\left.\vphantom{\hat T}T_1(0)\right|_{H_w}=h\left.
\vphantom{\hat T}T_2(0)\right|_{H_w}\label{17}\ee
holds, with $h$ a numeric factor.
\end{theorem}

{\it Proof.} According to the definition of $H_w$ and assumptions of the
theorem, we have
$$\left.\vphantom{\hat T}T_1(\lambda)T_3(\lambda)\right|_{H_w}=
h_1(\lambda),\quad
\left.\vphantom{\hat T}T_2(\lambda)T_3(\lambda)\right|_{H_w}=
h_2(\lambda),$$
where $h_1(\lambda)$, $h_2(\lambda)$ are functions such that $h_1(0)
\not=0$, $h_2(0)\not=0$. Putting $h=h_2(0)/h_1(0)$, we come to~(\ref{17}).
The theorem is proved.

\subsection{}\label{sec4}

The matrices $M_{\cal L}(\lambda)$ of vacuum curve coefficients of monodromy
matrices ${\cal L}(\lambda)$ introduced in Subsection~\ref{sec1} and
determined up to equivalence
$$M_{\cal L}(\lambda)\sim g(\lambda)M_{\cal L}(\lambda),\quad
g(\lambda)\not\equiv0,$$
form a group which we will denote $\cal G$. The composition law in that group
is consistent with the composition of $\cal L$-matrices (in the sense of
making monodromy matrices, as in~(\ref{4})), with $M_{\hat{\cal L}}(\lambda)$
being the inverse for $M_{\cal L}(\lambda)$ (Subsection~\ref{sec2}).

Define now for the subspace $H_w\subset H$ introduced in
Subsection~\ref{sec3} a subgroup ${\cal G}_w\subset{\cal G}$ that acts
projectively in $H_w$ in a natural way. Namely, ${\cal G}_w$ consists of the
matrices $M_{\cal L}(\lambda)$ for those ${\cal L}(\lambda)$ for which
$\det\left.\vphantom{\hat T}T(0)\right|_{H_w}\not=0$ and, as in
Theorem~\ref{t3}, a ${\cal L}_3(\lambda)$ exists such that
$M_{{\cal LL}_3}(\lambda)\equiv\left(\ba{cc}1&0\\0&1\ea\right)$
(equalities of this kind are understood, of course, to within a scalar
factor) and $\det T_3(0)\not=0$.

Then the action of ${\cal G}_w$ is given by the formula
\be M_{\cal L}\mapsto\left.\vphantom{\hat T} T(0)\right|_{H_w},\label{18}\ee
which is well-defined according to Theorem~\ref{t3}.

\subsection{}\label{sec5}

Thus, in this section a homomorphism was constructed from the semigroup of
monodromy matrices (with making of them ``larger'' monodromy matrices as
composition law) to a group of meromorphic $2\times2$-matrices depending on
$\lambda$ trigonometrically and determined up to a meromorphic scalar factor.
This homomorphism can be in a sense inverted (Subsection~\ref{sec4},
formula~(\ref{18})). The usefulness of this homomorphism will be shown in the
next section.

The constructions of this section can be extended to the case of even
$n=2p$ using ideas of~\cite{Kor-vac,Kor-vac-LOMI}. In particular, when
constructing monodromy matrices~(\ref{4}) one should use, instead of
$\cal L$-matrices~(\ref{6}--\ref{11}), the matrices ${\cal L}_+(\lambda)$
introduced in~\cite{Kor-vac,Kor-vac-LOMI}.

\section{Degeneracies in the spectrum of the 6-vertex model transfer matrix}
\label{realsec2}

\subsection{}\label{secs1}

Let $\eta=m\pi/n$, as in Section~\ref{realsec1}, with $m$ and $n$ relatively
prime integers. For simplicity, let us again, up to
Subsection~\ref{secs8}, assume that $n$ is odd. Denote as
$${\cal L}_0(\lambda)=\left(\ba{cc}{\cal A}_0(\lambda)&{\cal B}_0(\lambda)\\
{\cal C}_0(\lambda)&{\cal D}_0(\lambda)\ea\right)$$
the $(n-1)$th symmetric degree of the 6-vertex model $\cal L$-operator.
To be exact, ${\cal L}_0(\lambda)$ is such an $\cal L$-operator that
${\cal A}_0(\lambda),\ldots, {\cal D}_0(\lambda)$ act in a linear space of
dimension $n$ and possess a generating vector $\Omega$ with properties
$${\cal C}_0(\lambda)\Omega\equiv0,\quad{\cal A}_0(\lambda)\Omega=\sin\lambda
\cdot\Omega,\quad
{\cal D}_0(\lambda)\Omega=\sin(\lambda+2\eta)\cdot\Omega,$$
obtained from~(\ref{7}--\ref{10}) when $a=d=1,\;\rho=\sigma=
(1-n)\eta,\;b_{1n}=c_{n1}=0.$

Let
$$T_0(\lambda)={\rm Tr}\stackrel{
\makebox(0,0)[lb]{\put(0,0){\vector(-3,-2){0}}}%
\mbox{\normalsize\hspace{0.03em}\raisebox{-0.3ex}{$\frown$}}
}{
\prod_{i=1}^N
}
{\cal L}_0(\mu_i+\lambda)$$
be an inhomogeneous transfer matrix acting in the space $H$---the tensor
product of $N$ two-dimensional spaces (as in Subsection~\ref{sec3}).

Recall that $H_w$ denotes a common eigenspace of all transfer matrices
\be T(\lambda)={\rm Tr}\stackrel{
\makebox(0,0)[lb]{\put(0,0){\vector(-3,-2){0}}}%
\mbox{\normalsize\hspace{0.03em}\raisebox{-0.3ex}{$\frown$}}
}{
\prod_{i=1}^N
}
{\cal L}(\mu_i+\lambda)\label{s1}\ee
with the identity (strictly speaking, scalar) matrix of vacuum curve
coefficients
$$M_{\cal L}(\lambda)=\left(\ba{cc}1&0\\0&1\ea\right),$$
in particular, of transfer matrix $T_0(\lambda)$. The aim of this section is
to prove the following theorem.

\begin{theorem}\label{ts1} Let $w_0(\lambda)$ be an eigenvalue of transfer
matrix $T_0(\lambda)$ in $H_w$. If there are $K_w$ mutually different
${\rm mod}\, \pi/n$ zeroes $\lambda=\nu_1,\ldots,\nu_{K_w}$ among the {\em
simple\/} zeroes of the function $w_0(\lambda)$ (multiple zeroes are not
taken into account here) then $\dim H_w$ is divisible by $2^{K_w}$.
\end{theorem}

\subsection{}\label{secs2}

Let
\be{\cal L}(\lambda)=\left(\ba{cc}{\cal A}(\lambda)&{\cal B}(\lambda)\\
{\cal C}(\lambda)&{\cal D}(\lambda)\ea\right)\label{s2}\ee
be an $\cal L$-operator of the form~(\ref{6}--\ref{11}), i.e.\
${\cal A}(\lambda),\ldots,{\cal D}(\lambda)$ act in an $n$-dimensional space
while the generating vector $\Omega$ may not exist. Let $T(\lambda)$ be
given by formula~(\ref{s1}) and $\hat T(\lambda)$ be composed in the same way
 from the $\cal L$-operator~(\ref{s3a})
$$\hat{\cal L}(\lambda)=\left(\ba{cc}{\cal D}(\lambda)^T&-{\cal B}
(\lambda)^T\\
-{\cal C}(\lambda)^T&{\cal A}(\lambda)^T\ea\right).$$

\begin{lemma}\label{ls1}
In the previous paragraph notations,
\be T(\lambda)\hat T(\lambda)={\rm const}\cdot T_0(\lambda-\phi_1)
T_0(\lambda-\phi_2),\label{s4}\ee
where $\phi_1$ and $\phi_2$ are zeroes of the function $\det M_{\cal L}
(\lambda)$---the determinant of the vacuum curve coefficient matrix of the
operator ${\cal L}(\lambda)$~.
\end{lemma}

{\it Proof.} The statement that the formula~(\ref{s4}) is valid with some
$\phi_1$ and $\phi_2$ is a reformulation of lemmas~5 and 6 of~\cite{Kor-vac}
(see also~\cite{Kor-vac-LOMI}), and, according to the proof of the second of
those lemmas, $\phi_1$ and $\phi_2$ are zeroes of $\det {\cal L}(\lambda)$
(of multiplicity $n$). The fact that $\phi_1$ and $\phi_2$ are zeroes of
$\det M_{\cal L}(\lambda)$ (generally, of multiplicity one), easily follows
 from the explicit form of ${\cal L}(\lambda)$ and the definition
of~$M_{\cal L}(\lambda)$ (see Theorem~\ref{t1} and formula~(\ref{13})~). The
lemma is proved.

\subsection{}\label{secs4}

Let, besides the mentioned in Theorem~\ref{ts1} zeroes $\lambda=\nu_1,\ldots,
\nu_{K_w}$, the transfer matrix $T_0(\lambda)$ have in $H_w$ zeroes of
multiplicity $\geq\! 2$: \ $\lambda=\nu_{(K_w+1)}$, $\ldots\;$, $\nu_{M_w}
({\rm mod}\,\pi) $.
Let $T(\lambda)$ be a transfer matrix corresponding according to~(\ref{s1})
to such an operator (monodromy matrix) ${\cal L}(\lambda)$ whose vacuum curve
coefficient matrix $ M_{\cal L}(\lambda)$ is degenerate in the points
$\lambda =\phi_1,\ldots, \phi_q ({\rm mod}\,\pi/n)$: \ $\det M_{\cal L}
(\phi_i)=0,\quad 1\leq i \leq q$.

Recall that $ M_{\cal L}(\lambda)$ is defined up to a meromorphic scalar
factor $g(\lambda)$. We can thus assume that in each point $\phi_i$ the
entries of matrix $ M_{\cal L} $ are finite and not all equal to zero. The
fact that zeroes of $ \det M_{\cal L}(\lambda)$ are situated with period
$\pi/n$ follows from periodicity of $ M_{\cal L}(\lambda)$, see a remark in
the end of Subsection~\ref{sec1}.

\begin{lemma}\label{ls2}
If ${\cal L}(\lambda)$ and $T(\lambda)$ described above are such that for
all $i,j,\quad 1\leq i \leq q,\quad 1\leq j \leq M_w$,
\be \phi_i + \nu_j\not= 0 ({\rm mod}\, \pi/n),\label{s5} \ee
then there exists an $\cal L$-operator $\tilde{ \cal L} (\lambda)$
with the same vacuum curve
$$ M_{\tilde {\cal L}}(\lambda)=M_{\cal L}(\lambda) $$
such that the transfer matrix $\tilde T(\lambda)$ corresponding to it
according to~(\ref{s1}) is non-degenerate in $H_w$ for $\lambda=0$.
\end{lemma}

{\it Proof.} Let ${\cal L}_1(\lambda)$ be an $\cal L$-operator of the type
described in the beginning of Subsection~\ref{secs2}, and let us choose it so
that zeroes of $ {\rm det} M_{{\cal L}_1}(\lambda)$ be exactly in the points
$\phi_1$ and $\phi_2$ and the relation
\be{\rm Ker}\, M_{{\cal L}_1}(\phi_i) = {\rm Ker}\, M_{\cal L}(\phi_i), \quad
i=1,2 \label{s6} \ee
be valid (this can always be done, with changing, if necessary, the numbering
of points $\phi_i, \quad  1\leq i \leq q $). From~(\ref{s6}) it follows that
there exists a decomposition
$$ M_{\cal L}(\lambda)=M_{{\cal L}'}(\lambda) M_{{\cal L}_1}(\lambda),$$
where ${\rm det}M_{{\cal L}'}(\lambda)$ has by 2 zeroes ${\rm mod}\,\pi/n$
less than ${\rm det}M_{\cal L}(\lambda)$.
Proceeding further this way, we get
$$ M_{\cal L}(\lambda)=M_0M_{{\cal L}_{q/2}}(\lambda)
\ldots M_{{\cal L}_1}(\lambda),$$
where $M_0$ is a constant matrix, namely $M_0=\left(\ba{cc}a_0&0\\
0&d_0\ea \right)$ or $\left(\ba{cc}0&b_0\\c_0&0 \ea \right)$,
while ${\cal L}_2 (\lambda), \ldots , {\cal L}_{q/2}(\lambda)$ are
${\cal L}$-operators of the same type as ${\cal L}_1 (\lambda)$ (the total
number of zeroes $\phi_i$, counted with regard to their multiplicities,
is of course always even).

It follows from Lemma~\ref{ls1} that the transfer matrices $T_1(\lambda),
\ldots, T_{q/2}(\lambda)$ built of operators
${\cal L}_1(\lambda), \ldots,{\cal L}_{q/2}(\lambda)$
can be degenerate in $\lambda=0$ only if~(\ref{s5}) is violated. The same
applies, therefore, to the transfer matrix $\tilde T(\lambda)$ constructed
 from the $\cal L$-operator
$$\tilde{\cal L}={\cal L}_0{\cal L}_{q/2}(\lambda)\ldots{\cal L}_1(\lambda),$$
where ${\cal L}_0$, of course, corresponds to the matrix $M_0$.
Lemma~\ref{ls2} is proved.

\subsection{}\label{secs5}

Let now $T'(\lambda)$ and $T''(\lambda)$ be two transfer matrices constructed
according to formula~(\ref{s1})
 from operators ${\cal L}'(\lambda)$ and ${\cal L}''(\lambda)$
such that ${\rm det}\,M_{{\cal L}'}(\lambda)=0$
in the points $\phi'_1 ,\ldots, \phi'_{q'}$, and
${\rm det}\,M_{{\cal L}''}(\lambda)=0$ in the points
$\phi''_1 ,\ldots, \phi''_{q''}$. Let the conditions
$$\phi'_i + \nu_j \not =0 \quad ({\rm mod}\,\pi/n),$$
$$ \phi''_i + \nu_j \not =0 \quad ({\rm mod}\,\pi/n),$$
where, as before, $\nu_j, \quad 1\leq j \leq M_w$, are zeroes of
$\left.\vphantom{\hat T}T_0(\lambda)\right|_{H_w}$,
be valid for all $i,j$, except $i=j=1$, i.e.\
$$\phi'_1+\nu_1=0,$$
$$\phi'_1=\phi''_1.$$

\begin{lemma}\label{ls3}
If the operators ${\cal L}'(\lambda)$ and ${\cal L}''(\lambda)$ from the
previous paragraph are such that
\be{\rm Ker}\, M_{{\cal L}'}(-\nu_1) = {\rm Ker}\, M_{{\cal L}''}(-\nu_1),
\label {s7} \ee
then
\be{\rm Ker}\,\left.\vphantom{\hat T}T_0'(0)\right|_{H_w}={\rm Ker}\,
\left.\vphantom{\hat T}T_0''(0)\right|_{H_w}\label {s8}. \ee
\end{lemma}

{\it Proof.} One can find in the same way as in the proof of
Lemma~\ref{ls1} that
$$T'(\lambda)\hat T'(\lambda)={\rm const} \cdot \prod^{q'}_{i=1}
T_0(\lambda-\phi'_i).$$
Thus, $\left.T'(\lambda)\hat T'(\lambda)\right|_{H_w}$
is a scalar operator having a simple zero in $\lambda=0$, whence
\be\left.\vphantom{\hat T}{\rm Ker}\,T'(0)\right|_{H_w} ={\rm Im}\,
\left.\hat T'(0)\right|_{H_w}\label{s9}. \ee
 From~(\ref{s7}) and the fact that $M_{\hat{\cal L}'}(\lambda)$ is
proportional to $(M_{{\cal L}'}(\lambda))^{-1}$ it follows that
$M_{{\cal L}''}(\lambda) M_{\hat{\cal L}'}(\lambda)$
is non-degenerate in $\lambda=-\nu_1$ (to within a scalar factor!).
Thus, ${\cal L}(\lambda)={\cal L}''(\lambda) \hat{\cal L}'(\lambda)$
satisfies the conditions of Lemma~\ref{ls2}, from which it follows that at
$\lambda \rightarrow 0 \quad
\left.T''(\lambda)\hat T'(\lambda)\right|_{H_w}$
is proportional to a non-degenerate operator (the operator~$\tilde T(0)$ in
Lemma~\ref{ls2} notations). Thus,
\be \left.\vphantom{\hat T}{\rm Ker}\,T_0''(\lambda)\right|_{H_w} =\left.
{\rm Im}\,\hat T'(0)
\right|_{H_w} \label {s10} \ee
Comparing~(\ref{s9}) with~(\ref{s10}), we come to~(\ref{s8}). The lemma is
proved.

\subsection{}\label{secs6}

Now in this subsection let the operator
$${\cal L}(\lambda)=\left(\ba{cc}{\cal A}(\lambda)&{\cal B}(\lambda)\\
{\cal C}(\lambda)&{\cal D}(\lambda)\ea\right)$$
and the transfer matrix $T(\lambda)$ constructed according to~(\ref{s1})
have the following properties:

a) ${\cal A}(\lambda), \ldots ,{\cal D}(\lambda)$ act in an
$n$-dimensional space, with no generating vector $\Omega$ possessing the
property ${\cal C}\Omega\equiv 0$;

b) the vacuum curve coefficient matrix has the form
$$M_{\cal L}(\lambda)=\left(\ba{cc}\alpha(\lambda)&\beta(\lambda)\\
\beta(\lambda)&\alpha(\lambda)\ea\right);$$

c)
$$T(\lambda)\hat T(\lambda)={\rm const} \cdot T_0(\lambda-\phi_1)
T_0(\lambda-\phi_2),$$
with all the sums
$$\phi_i+\nu_j, \quad i=1,2; \quad  1\leq j \leq M_w$$
pairwise different (recall that $\nu_j$ are zeroes of the scalar operator
$T_0(\lambda)\left.\vphantom{\hat T}\right|_{H_w}$,
and the first $K_w$ of them are the simple ones).

${\cal L}(\lambda)$ with properties a)--c) always exists. It follows from the
property b) that $M_{\cal L}(\lambda)$ and $M_{\cal L}(\mu)$ commute for all
$\lambda,\mu$, so $T(\lambda)$ and $T(\mu)$, and also $T(\lambda)$ and
$\hat T(\mu)$, commute as well.

\begin{lemma}\label{ls4}
For $1\leq j \leq K_w$ the decomposition takes places
$$H_w=\left.\vphantom{\hat T}{\rm Ker }\,T(\phi_1 + \nu_j) \right|_{H_w}
\oplus\left.{\rm Ker }\,\hat T(\phi_1 + \nu_j) \right|_{H_w}. $$
\end{lemma}

{\it Proof.} The scalar operator
$$ \left.\left(T(\phi_1 + \nu_j+
\lambda)\hat T(\phi_1 + \nu_j+\lambda) \right) \right|_{H_w}$$
has a simple zero in $\lambda=0$, which can be written in the form
\begin{eqnarray}
\biggl(
T(\phi_1 + \nu_j) \hat T(\phi_1 + \nu_j) + \left.
\lambda \frac{dT(\phi_1+\nu_j+\lambda)}{d\lambda} \right|_{\lambda=0}
\hat T(\phi_1 + \nu_j)+  \nonumber\\
+\left.\left.\lambda T(\phi_1 + \nu_j) \frac{d\hat T(\phi_1+\nu_j+\lambda)}
{d\lambda} \right|_{\lambda= 0}\;
\biggr)
\right|_{H_w}={\rm const}\cdot\lambda +o(\lambda). \nonumber
\end{eqnarray}
 From the terms of order zero in $\lambda$ we get
\be\left.\dim {\rm Ker}\,\vphantom{\hat T}T(\phi_1+\nu_j)\right|_{H_w}+
\left.\dim {\rm Ker}\,\hat T(\phi_1+\nu_j)\right|_{H_w}\geq
\dim H_w.\label{s11}\ee
 From the terms of the first order in $\lambda$ we see, taking into account
the commutativity of $T$ and $d\hat T/d\lambda$ which follows, as was
explained above, from the condition b) of this subsection, that there cannot
exist a nonzero  vector $\Phi\in H_w$ with properties
$T(\phi_1+\nu_j)\Phi=0$ and $\hat T(\phi_1+\nu_j)\Phi=0$, which means
\be\left.{\rm Ker}\,\vphantom{\hat T}T(\phi_1+\nu_j)\right|_{H_w}\cap
\left.{\rm Ker}\,\hat T(\phi_1+\nu_j)\right|_{H_w}=0.\label{s12}\ee
Relations~(\ref{s11}) and~(\ref{s12}) together mean exactly what was required
in the lemma, so the proof is complete.

For each subset $A\subset\{1,\ldots,K_w\}$ of the set of integers from
1 to $K_w$ let us introduce a subspace $H(A)\subset H_w$:
$$H(A)=\bigcap_{i\in A}\left.\vphantom{\hat T}{\rm Ker}\,T(\phi_1+\nu_i)
\right|_{H_w}
\bigcap_{\mbox{\scriptsize$\ba{l}j=1\\j\not\in A\ea$}}^{K_w}\left.
{\rm Ker}\,\hat T(\phi_1+\nu_j)\right|_{H_w}.$$

\begin{lemma}\label{ls5}
The dimensions if subspaces $H(A)$ are equal for all $A$; there is a
decomposition
\be H_w=\bigoplus_A H(A).\label{s13}\ee
\end{lemma}

{\it Proof.} The decomposition~(\ref{s13}) readily follows from
Lemma~\ref{ls4} and the commutativity of $T(\lambda),T(\mu),
\hat T(\lambda')$ and $\hat T(\mu')$ for all $\lambda,\mu,\lambda',\mu'$. To
prove the equalness of the dimensions of $H(A)$, it is sufficient to
construct for any pair $A_1,A_2$ a non-degenerate operator $F$ mapping
$H(A_1)$ into $H(A_2)$. Let e.g.\ $A_1=\{1,\ldots,K_w\}$ and $A_2=
\{2,\ldots,K_w\}$. Let us construct the operator $F$ with properties
$$\ba{l}
\left.F\,{\rm Ker}\,\vphantom{\hat T}T(\phi_1+\nu_1)\right|_{H_w}=
\left.F\,{\rm Ker}\,\hat T(\phi_1+\nu_1)\right|_{H_w},\smallskip\\
\left.F\,{\rm Ker}\,\vphantom{\hat T}T(\phi_1+\nu_2)\right|_{H_w}=
\left.F\,{\rm Ker}\,\vphantom{\hat T} T(\phi_1+\nu_2)\right|_{H_w},
\smallskip\\
\dotfill\smallskip\\
\left.F\,{\rm Ker}\,\vphantom{\hat T}T(\phi_1+\nu_{K_w})\right|_{H_w}=
\left.F\,{\rm Ker}\,\vphantom{\hat T}T(\phi_1+\nu_{K_w})\right|_{H_w}.\ea$$
Applying Lemma~\ref{ls3} we find that we can set
$$F=\left.\tilde T(0)\right|_{H_w},$$
where $\tilde T(\lambda)$ is the transfer matrix built of an operator
$\tilde{\cal L}(\lambda)$ with properties
\be\left.\ba{l}
M_{\tilde{\cal L}}(-\nu_1){\rm Ker}\,M_{\cal L}(\phi_1)={\rm Ker}\,
M_{\hat{\cal L}}(\phi_1),\smallskip\\
M_{\tilde{\cal L}}(-\nu_2){\rm Ker}\,M_{\cal L}(\phi_1)={\rm Ker}\,
M_{\cal L}(\phi_1),\smallskip\\
\dotfill\smallskip\\
M_{\tilde{\cal L}}(-\nu_{K_w}){\rm Ker}\,M_{\cal L}(\phi_1)={\rm Ker}\,
M_{\cal L}(\phi_1).\ea\right\}\label{s14}\ee
Recall that $M_{\tilde{\cal L}}(\lambda)$ consists of trigonometrical
polynoms whose degree depends on $\tilde{\cal L}$. Choosing this degree big
enough, one can satisfy all the conditions~(\ref{s14}) together with
nondegeneracy of $\left.\tilde T(0)\right|_{H_w}$. The lemma is proved.

The proof of Theorem~\ref{ts1} comes now to its end with an observation that
the number of subspaces $H(A)$ equals $2^{K_w}$.

\subsection{}\label{secs7}

Let us apply the obtained results to calculating the degeneracy multiplicity
of the 6-vertex model transfer matrix eigenvalue corresponding to the ``naked
vacuum'', i.e.\ the eigenvector
$$\left(\ba{c}1\\0\ea\right)\otimes\ldots\otimes\left(\ba{c}1\\0\ea\right).$$
Let us assume that the chain length $N$ is a multiple of $n$. A simple
calculation shows that in this case $K_w=N/n$. Hence, the degeneracy
multiplicity is divisible by $2^{N/n}$.

\subsection{}\label{secs8}

Thus, the results of Section~\ref{realsec1} have been applied to calculating
the degeneracy multiplicities of the 6-vertex model transfer matrix spectrum.
These multiplicities turned out to be divisible by high (as it is seen from
the example in Subsection~\ref{secs7}) degrees of the number 2.

In the case of even $n=2p$, one can perform all the reasoning in much the
same way as above. Some necessary complications follow from the
paper~\cite{Kor-vac} (or~\cite{Kor-vac-LOMI}). In particular, the transfer
matrices must be constructed using the operator ${\cal L}_+(\lambda)$
(\cite{Kor-vac}, formula~(30)) instead of~${\cal L}(\lambda)$.
Theorem~\ref{ts1} remains valid for $n=2p$ if one changes ${\rm mod}\,\pi/n$
to ${\rm mod}\,\pi/p$ in its formulation.

\end{document}